# A Distributed and Deterministic TDMA Algorithm for Write-All-With-Collision Model[*]

Mahesh Arumugam

Cisco Systems, Inc.,
San Jose, CA 95134
Email: `maarumug@cisco.com`

**Abstract**

Several self-stabilizing time division multiple access (TDMA) algorithms are proposed for sensor networks. In addition to providing a collision-free communication service, such algorithms enable the transformation of programs written in abstract models considered in distributed computing literature into a model consistent with sensor networks, i.e., write all with collision (WAC) model. Existing TDMA slot assignment algorithms have one or more of the following properties: (i) compute slots using a randomized algorithm, (ii) assume that the topology is known upfront, and/or (iii) assign slots sequentially. If these algorithms are used to transform abstract programs into programs in WAC model then the transformed programs are probabilistically correct, do not allow the addition of new nodes, and/or converge in a sequential fashion. In this paper, we propose a self-stabilizing deterministic TDMA algorithm where a sensor is aware of only its neighbors. We show that the slots are assigned to the sensors in a concurrent fashion and starting from arbitrary initial states, the algorithm converges to states where collision-free communication among the sensors is restored. Moreover, this algorithm facilitates the transformation of abstract programs into programs in WAC model that are deterministically correct.

---

[*]Contact Information:
  Address: 170 W. Tasman Dr, San Jose, CA 95134
  Phone: +1-408-853-3547
  Fax: +1-408-527-9537
  URL: http://aumahesh.googlepages.com/

# 1 Introduction

One of the important concerns in programming distributed computing platforms is the model of computation used to specify programs. Programs written for distributed computing platforms such as sensor networks and embedded systems often have to deal with several low level challenges of the platform. In sensor networks, especially, one has to write programs that deal with issues such as communication, message collision and race conditions among different processes. Therefore, to simplify the programming, it is important to abstract such low level issues. In other words, the ability to specify programs in an abstract model and later transform them into a concrete model that is appropriate to the platform is crucial.

The problem of transformation of programs in an abstract model to programs in other models of computation has been studied extensively (e.g., [1–6]). These transformations cannot be applied to obtain concrete programs for sensor networks as the model of computation in sensor networks is *write all with collision* (WAC) model. In WAC model, communication is *local broadcast* in nature. As a result, whenever a sensor executes an *action*, it writes the state of all its neighbors in one atomic step. However, if two neighbors $j$ and $k$ of a sensor (say $i$) try to execute their write actions at the same time then, due to collision, state of $i$ will remain unchanged. The actions of $j$ and $k$ may update the state of their other neighbors successfully.

**Existing transformations for WAC model.** Recently, several approaches have been proposed to transform programs written in abstract models considered in distributed computing literature into programs in WAC model [7–10]. Such transformation algorithms can be classified into two categories: (a) randomized [7, 8] and (b) deterministic [9, 10].

In [7], the authors propose a *cached sensornet transform* (CST) that allows one to correctly simulate an abstract program in sensor networks. This transformation uses carrier sensor multiple access (CSMA) based MAC protocol to broadcast the state of a sensor and, hence, the transformed program is randomized. And, the algorithm in [9] uses time division multiple access (TDMA) to ensure that collisions do not occur during write actions. Specifically, in WAC model, each sensor executes the *enabled* actions in the TDMA slots assigned to that sensor. And, the sensor writes the state of all its neighbors in its TDMA slots. If the TDMA algorithm in [11], a self-stabilizing and deterministic algorithm designed for grid-based topologies, is used with [9] then the transformed program in WAC model is self-stabilizing and deterministically correct for grid-based topologies. And, if randomized TDMA algorithms proposed in [8, 12] are used with [9] then the transformed program is probabilistically correct. Finally, the algorithm in [10], a self-stabilizing and deterministic TDMA algorithm for arbitrary topologies, allows one to transform abstract programs into programs in WAC model that are deterministically correct for arbitrary topologies.

In this paper, we are interested in stabilization preserving deterministic transformation for WAC model. As mentioned above, a self-stabilizing deterministic TDMA algorithm enables such a transformation. One of the drawbacks of existing self-stabilizing deterministic TDMA algorithms (e.g., [10]) is that the recovery is sequential. Specifically, in [10], whenever the network is perturbed to states where the TDMA slots are not collision-free, a distinguished sensor (e.g., base station) initiates a recovery process and each sensor recomputes its slots one by one. However, it is desirable that the network self-stabilizes from such arbitrary states in a distributed and concurrent manner (without the assistance of distinguished sensors).

**Contributions of the paper.** To redress this deficiency, in this paper, we propose a self-stabilizing deterministic TDMA algorithm that provides concurrent recovery. In this algorithm, whenever a sensor observes that the slots assigned to its neighbors are not collision-free, it initiates a recovery. As a result, its neighbors recover to legitimate states (i.e., the slots are collision-free) and the network as a whole self-stabilizes concurrently. We show that the algorithm supports addition or removal of sensors in the network. While a removal of a sensor does not affect the normal operation of the network, our algorithm ensures that the slots assigned to removed sensor are reused. And, our algorithm supports *controlled* addition of new sensors in the network. In addition, we propose an extension to our algorithm that improves the bandwidth allocation of the sensors.

**Organization of the paper.** The rest of the paper is organized as follows. In Section 2, we introduce the models of computation considered in distributed computing platforms. We formally state the problem definition of TDMA and identify the assumptions made in this paper. In Section 3, we present our distributed TDMA slot assignment algorithm for WAC model. We show that the algorithm is self-stabilizing. Also, we discuss extensions that allow dynamic addition/removal of sensors, improve bandwidth allocation, and synchronize time. Then, in Section 4, we compare our algorithm with related work. Finally, in Section 5, we summarize our work and identify future research directions.

## 2 Preliminaries

In this section, we define the models of computation, formally state the problem, and discuss the assumptions made in this paper.

### 2.1 Models of Computation

A computation model limits the variables that a program can read and write. Program actions are split into a set of processes (i.e., sensors). Each action is associated with one of the processes in the program. We now define shared memory model and WAC model.

**Shared memory model.** In this model, in one atomic step, a sensor can read its state as well as the state of all its neighbors and write its own state.

**Write all with collision (WAC) model.** In this model, each sensor consists of write actions (to be precise, write-all actions). In one atomic step, a sensor can update its own state and the state of all its neighbors. However, if two or more sensors simultaneously try to update the state of a sensor, say, $k$, then the state of $k$ remains unchanged. Thus, WAC model captures the fact that a message sent by a sensor is broadcast. But, if multiple messages are sent to a sensor simultaneously then, due to collision, it receives none.

### 2.2 Problem Statement

**Distributed TDMA slot assignment.** TDMA is the problem of assigning communication time slots to each sensor. Two sensors $j$ and $k$ cannot transmit in the same slot if their communication interferes with each other. In other words, $j$ and $k$ cannot transmit in the same slot if the communication distance between them is less than or equal to 2. To model this requirement, we consider the sensor network as a graph $G = (V, E)$ where $V$ is the set of all sensors and $E$ is the communication topology of the network. More precisely, if sensors $u \in V$ and $v \in V$ can communicate with each other then the edge $(u, v) \in E$. Finally, $distance_G(u, v)$ identifies the communication distance between $u$ and $v$ in $G$. The communication distance is the number of links in the shortest path between the two sensors. Thus, the problem statement of TDMA is shown in Figure 1.

> **Problem Statement: Distributed TDMA Slot Assignment**
> Consider the communication graph $G=(V, E)$; Given a sensor $j \in V$, assign time slots to $j$ such that the following condition is satisfied:
> $k \in V \land k \neq j \land distance_G(j, k) \leq 2 \Longrightarrow slot.j \cap slot.k = \emptyset$
> where $slot.i$ identifies the slots assigned to sensor $i$.

Figure 1: Problem statement of distributed TDMA slot assignment

**Definition 2.1** *(TDMA frame)  In TDMA, time is partitioned into fixed sized frames. Each TDMA frame is divided into fixed sized slots. In this paper, we ensure uniform bandwidth allocation among sensors. Therefore, each sensor is assigned one slot in every TDMA frame. A sensor is allowed to transmit in the slots assigned to it.*

**Definition 2.2** *(TDMA period)  The length of the TDMA frame is called the TDMA period. More specifically, it is the interval between the slots assigned to a sensor in consecutive frames.*

**Distance 2 coloring.**   The problem statement of TDMA is similar to the problem of distance 2 coloring. Distance 2 coloring algorithm assigns colors to all the sensors in the network such that the colors assigned to distance 2 neighborhood of a sensor are unique. The color assigned to a sensor identifies the initial TDMA slot of that sensor. The sensor can compute its subsequent TDMA slots using TDMA period. Ideally, TDMA period $P = (d^2 + 1)$, where $d$ is the maximum degree of the network. (We refer the reader to [10] for a proof that the number of colors required to obtain distance 2 coloring is at most $d^2 + 1$.) Thus, Figure 2 states the problem definition of distance 2 coloring.

> **Problem Statement: Distance 2 Coloring**
> Consider the communication graph $G = (V, E)$; Given a sensor $j \in V$, assign a color to $j$ such that the following condition is satisfied:
> $k \in V \land k \neq j \land \text{distance}_G(j, k) \leq 2 \implies \text{color}.j \neq \text{color}.k$
>     where *color.i* identifies the color assigned to sensor $i$.

Figure 2: Problem statement of distance 2 coloring

**Self-stabilization.**   An algorithm is self-stabilizing iff starting from an arbitrary state, it: (a) recovers to legitimate state and (b) upon recovery continues to be in legitimate states forever [13,14]. Extending this definition, we have the problem statement of a self-stabilizing TDMA slot assignment algorithm as shown in Figure 3.

> **Problem Statement: Self-Stabilizing TDMA Slot Assignment**
> Consider the communication graph $G = (V, E)$; A TDMA slot assignment algorithm is self-stabilizing iff starting from arbitrary initial states, the algorithm recovers to the following state:
> $j \in V \land k \in V \land k \neq j \land \text{distance}_G(j, k) \leq 2 \implies \text{slot}.j \cap \text{slot}.k = \emptyset$
> and continues to remain in this state forever.

Figure 3: Problem statement of self-stabilizing TDMA slot assignment

## 2.3  Assumptions

In this paper, we do not assume the presence of a base station. In our algorithm, the sensors collaborate among themselves to obtain distance 2 coloring and TDMA slot assignments. We assume that each sensor knows the IDs of the sensors that it can communicate with. This assumption is reasonable since the sensors collaborate among their neighbors when an event occurs. We assume that the maximum degree of the graph does not exceed a certain threshold, say $d$. This can be ensured by having the deployment follow a certain geometric distribution or using a predetermined topology. Finally, we assume that the clocks of the sensors are synchronized. Later, in Section 3.5, we discuss how sensors can synchronize their clocks.

# 3 TDMA Slot Assignment Algorithm

In this section, we present our distributed and deterministic TDMA algorithm. First, in Section 3.1, we give the outline of the algorithm. Then, in Section 3.2, we present the algorithm in detail. Specifically, we discuss how the network self-stabilizes starting from arbitrary initial states to states where the slots are assigned as identified in Figure 3. Subsequently, in Section 3.3, we illustrate our algorithm with an example. And, in Section 3.4, we discuss the convergence and scalability properties of the algorithm. Finally, in Section 3.5, we discuss extensions to this algorithm.

## 3.1 Outline of the Algorithm

Initially, the colors assigned to the sensors may be arbitrary. As a result, the communication among the sensors may not be collision-free. To achieve collision-free communication among the sensors, in this algorithm, we adopt *distributed reset* (e.g., [15]) approach. More specifically, whenever collisions are observed for a particular slot (i.e., color) for a threshold number of consecutive TDMA frames (say, at $j$), the algorithm resets the colors of appropriate sensor(s) in the neighborhood of $j$. In other words, a reset computation is used to update the colors assigned to the sensors such that the sensors in distance 2 neighborhood of $j$ have unique colors and, thus, ensure that slots assigned to them are collision-free at $j$.

Towards this end, $j$ schedules a reset computation in its current TDMA slots. It schedules the reset such that the following requirements are satisfied: (i) reset computations of others sensors in the distance 2 neighborhood of $j$ do not interfere with each other and (ii) when $j$ initiates reset, the sensors in the distance 3 neighborhood of $j$ have stopped transmitting. The first requirement ensures that only one reset computation is active in a given neighborhood at any instant. Otherwise, simultaneous resets in a distance 2 neighborhood may result in collisions and/or sensors choosing conflicting colors. The second requirement ensures that the reset messages and update messages are communicated in a collision-free manner.

Whenever a sensor, say $k$, receives the reset message from $j$, first, it updates the color information it maintains about its distance 1 and distance 2 neighbors. Next, it checks if it has to change the color in response to the reset. If $k$ needs to update its color, it chooses a non-conflicting color among the sensors in its distance 2 neighborhood. And, subsequently, $k$ broadcasts change color message in its newly computed slots.

Now, whenever a sensor, say $l$, receives the change color message from $k$, first, it cancels any scheduled reset computations. Subsequently, $l$ updates the color information it maintains about its distance 1 and distance 2 neighbors. When $j$ receives change color message, it sends restart message to signal its distance 3 neighborhood to restart application communication. Thus, the algorithm resets the neighborhood of $j$ to deal with a collision at $j$. However, note that one reset computation may not be sufficient to restore the state of the network.

## 3.2 Reset Computation and Slot/Color Assignment

In this section, we discuss the algorithm in detail. This is a 5-step algorithm: (1) observe collision and schedule reset computation, (2) send reset message, (3) update color, (4) notify color, and (5) restart communication. Now, we discuss each of these 5 steps. These steps may be repeated until the network self-stabilizes to legitimate states.

**Step 1: Observe collision and schedule reset computation.** If a sensor, say $j$, observes collision at slot $c_x$ (i.e., color $c_x$) for a threshold number of consecutive frames then it schedules a reset computation. Towards this end, first, $j$ appends $c_x$ to *collisions.j*, the list of collision slots it has observed so far. Also, it adds $(f_c.j, c_x)$ to *timestamp.j*, where $f_c.j$ is the frame in which $j$ observed the collision at slot $c_x$. If $j$ observed a collision for the first time then $j$ determines the slot in which it can send a reset message. Sensor $j$ schedules a reset computation such that requirements identified in Section 3.1 are met.

*Requirement 1: Ensure only one active reset in the neighborhood.* To satisfy this requirement, $j$ schedules the reset computation in TDMA frame $f_{reset}.j = f_c.j + ID.j + D3_{timeout}$, where $ID.j$ is the ID of sensor $j$ and $D3_{timeout}$ is defined below. This ensures that if two sensors observe a collision simultaneously, then their resets are scheduled in unique frames. On the other hand, if the sensors observe a collision in different frames, it is possible that their resets are scheduled in the same frame. However, before a sensor initiates a reset, requirement 2 ensures that the distance 3 neighborhood has stopped. As a result, the sensor that observed a collision earlier will be able to proceed with the reset without any collision.

*Requirement 2: Ensure distance 3 neighborhood has stopped.* Suppose $j$ has scheduled reset in $f_{reset}.j$. Before $j$ initiates reset, it has to wait until its distance 3 neighborhood stops transmitting messages. Towards this end, $j$ stops transmitting for *at least* $D3_{timeout}$ frames before it fires the reset. $D3_{timeout}$ is the number of TDMA frames required for distance 3 neighborhood of $j$ to stop transmitting messages. Specifically, when $j$ stops, its neighbors will notice that $j$ has stopped. As a result, distance 1 neighbors of $j$ stop. Likewise, distance 2 and distance 3 neighbors of $j$ stop. To prevent false positives, neighbor, say $l \in N.j$, stops only after it detects that $j$ has stopped for a threshold number of consecutive frames, $stop_{timeout}$. Therefore, in order to ensure that distance 3 neighborhood of a sensor has stopped, $D3_{timeout} \geq 3 stop_{timeout}$.

**Step 2: Send reset message.** Each sensor, say $j$, maintains the state of its distance 2 neighborhood: $nbrClr.j$ (contains the state of distance 1 neighbors of $j$) and $dist2Clr.j$ (contains the state of distance 2 neighbors of $j$). Each entry in $nbrClr.j$ contains color assignment and the last frame in which $j$ or its neighbors received a message from the corresponding sensor. Likewise, each entry in $dist2Clr.j$ contains color assignment and the last frame in which one of the neighbors of $j$ received a message from the corresponding sensor. Initially, $nbrClr.j$ contains arbitrary color assignments that may not reflect the accurate state of all its distance 1 neighbors. And, $dist2Clr.j$ may not reflect the actual distance 2 neighbors and their state.

**Notation.** We denote an entry in $nbrClr.j$ as $(k, c_k, f_k)$; this indicates that $j$ last received a message from $k$ in frame $f_k$ and in slot (i.e., color) $c_k$. Entries in $dist2Clr.j$ are denoted similarly. Additionally, we use "-" to wildcard or *ignore* a field in an entry. For example, $(-, c_x, -)$ indicates that we are interested in entries that have the color $c_x$. Additionally, we denote the current frame at $j$ as $f_{current}.j$.

Sensor $j$ initiates a reset in frame $f_{reset}.j$ only if it has not stopped transmitting in response to another reset. From Step 1, we note that $j$ sends the reset message to its distance 1 neighbors in a collision-free manner. The reset message format is shown in Figure 4. The message includes the state of distance 1 neighbors that $j$ knows currently, list of collisions and their timestamps, the sensor that should update its color in response to this reset, and the initiator of the reset (in this case, $j$). Sensor $j$ selects the sensor that should update its color based on IDs of the neighbors that $j$ did not hear for a threshold number of consecutive frames.

| | neighbor | color | lastReceived |
|---|---|---|---|
| $rm_j.neighborState$ | $j$ | $color.j$ | $f_{current}.j$ |
| | $nbrClr.j$ | | |
| $rm_j.collisionInfo$ | $collisions.j$ | | |
| $rm_j.resetTimestamp$ | $timestamp.j$ | | |
| $rm_j.sensorToChange$ | $l$, where $l \in N.j$ is the sensor with lowest ID for which $j$ did not hear any thing for a threshold number of frames | | |
| $rm_j.initiator$ | $j$ | | |

Figure 4: Reset message of $j$, $rm_j$

**Theorem 3.1** *Reset computation initiated by any sensor executes in a collision-free manner.*

**Proof.** Suppose two reset computations execute simultaneously in a distance 2 neighborhood. Let $k$ and $l$ be two unique sensors that have initiated the reset such that $distance_G(k,l) \leq 2$. Both $k$ and $l$ should have observed a collision in the same frame and scheduled resets to start at the same frame. Otherwise, either one of them would have observed that that the neighbors have stopped in response to a reset of the other and, hence, it would have stopped as well. Therefore, we have, $f_{reset}.k = f_{reset}.l$. In other words, $f_c.k + ID.k + D3_{timeout} = f_c.l + ID.l + D3_{timeout}$. Without loss of generality, assume that $ID.k < ID.l$. Now, we have $f_c.k > f_c.l$. More specifically, $l$ observed the collision before $k$ did. This is a contradiction. □

**Step 3: Update color, if necessary.** Whenever a sensor, say $k$, receives the reset message $rm_j$, first, it cancels any scheduled reset. Next, it updates its neighbor state using the information in $rm_j$. Specifically, it updates $nbrClr.k$ with the color information of the initiator of the reset $j$. And, it updates $nbrClr.k$ and $dist2Clr.k$ using the information in $rm_j$ about distance 1 neighbors of $j$. Towards this end, we proceed as shown in Figure 5. (Note that $k$ updates an entry in $nbrClr.k$ or $dist2Clr.k$ only if the initiator $j$ had received a message from the corresponding sensor most recently than that of $k$.)

---

if $(j = rm_j.initiator \land (j, c_j, -) \in rm_j.neighborState)$
   $nbrClr.k = \{nbrClr.k - (j, -, -)\} \cup (j, c_j, f_{current}.k)$
if $(p \in N.j \land (p, c_p, f_1) \in rm_j.neighborState \land (p, -, f_2) \in nbrClr.k \land f_2 < f_1)$
   $nbrClr.k = \{nbrClr.k - (p, -, -)\} \cup (p, c_p, f_1)$
else if $(p \notin N.k \land (p, c_p, f_1) \in rm_j.neighborState \land (p, -, f_2) \in dist2Clr.k \land f_2 < f_1)$
   $dist2Clr.k = \{dist2Clr.k - (p, -, -)\} \cup (p, c_p, f_1)$

---

Figure 5: Updating $nbrClr.k$ and $dist2Clr.k$ of sensor $k$

Sensor $k$ then checks if it has to update its color. If $k = rm_j.sensorToChange$ then $j$ requires $k$ to update its color. Sensor $k$ updates its color as shown in Figure 6. Specifically, if $color.k$ is in $rm_j.collisionInfo$, $k$ chooses a color $c$ from $K$ (i.e., the set of all available colors) such that there is no collision in slot $c$ at $j$ and is unique among its distance 2 neighborhood.

---

if $(k = rm_j.sensorToChange \land color.k \in rm_j.collisionInfo)$ {
   $potentialColors = \{c | c \in K \land c \notin rm_j.collisionInfo \land (-, c, -) \notin nbrClr.k \land (-, c, -) \notin dist2Clr.k\}$
   $color.k = min(potentialColors)$
}

---

Figure 6: Updating color assignment of sensor $k$

**Step 4: Notify color.** If $k = rm_j.sensorToChange$, it sends *change color message* $cm_k$ to all its neighbors as shown in Figure 7 (regardless of whether it changed its color or not). Specifically, $k$ sends its color information, $nbrClr.k$, and the initiator of the reset. Whenever a sensor receives change color message, first, it cancels any scheduled resets. Next, it updates its $nbrClr$ and $dist2Clr$ similar to the discussion shown in Figure 5. Specifically, if $l$ receives $cm_k$, it updates $nbrClr.l$ with $(k, c_k, f_{current}.l)$, where $(k, c_k, -) \in cm_k.neighborState$. Similarly, $l$ updates $nbrClr.l$ and $dist2Clr.l$ based on neighbor state information in $cm_k$.

| | neighbor | color | lastReceived |
|---|---|---|---|
| $cm_k.neighborState$ | $k$ | $color.k$ | $f_{current}.k$ |
| | $nbrClr.k$ | | |
| $cm_k.initiator$ | $j$ | | |

Figure 7: Change color message of $k$, $cm_k$

**Theorem 3.2** *If a sensor updates its color in response to a reset then the change color message of that sensor is communicated in a collision-free manner.*

**Proof.** Let $j$ be the initiator of the reset. And, $l \in N.j$ updates its color in response to the reset of $j$. When $j$ initiates the reset ($rm_j$), distance 3 neighbors of $j$ have stopped transmitting. Therefore, when $l$ sends change color message $cm_l$, neighbors of $l$ will receive it successfully. Hence, all neighbors of $l$ will get the latest color assigned to $l$. □

**Step 5: Restart communication.** Whenever $j$ initiates a reset, it expects to receive a change color message from $rm_j.sensorToChange$ before its next allotted slot in $f_{current}.j + 1$ frame. If $j$ receives the change color message from the sensor that changed the color in response to reset of $j$, $j$ cleans $collisions.j$ and $timestamp.j$. Then, it signals its neighbors to restart application communication. Specifically, it sends restart message, $sm_j$; the format of $sm_j$ is the same as change color message as shown in Figure 7. Once a sensor receives $sm_j$, it updates $nbrClr$ and $dist2Clr$ and starts application communication in its slots. Continuing in this fashion, the distance 3 neighborhood of $j$ restart their communication. Note that the restart operation updates the color assignment of $l = rm_j.sensorToChange$ at distance 2 neighborhood of $l$, potentially causing collisions at some distance 2 neighbors of $l$. When a sensor hears a restart message or collision, it restarts application communication.

On the other hand, if $l = rm_j.sensorToChange$ did not send change color message (possibly, due to failure of $l$) then $j$ marks $l$ as *potentially failed*. And, it cleans $collisions.j$ and $timestamp.j$. Also, it sends the restart message as mentioned above. In future resets at $j$, $j$ will not set $l$ in $rm_j.sensorToChange$. If $l$ has indeed failed, the extension proposed in Section 3.5.1 will reclaim the slots assigned to $l$. Otherwise, sensor $j$ will remove $l$ from the list of potentially failed sensors when $j$ hears a message from $l$.

**Theorem 3.3** *If a sensor changes its color in response to a reset, eventually, the distance 2 neighborhood of that sensor learn the state of the sensor.*

**Proof.** Suppose $k \in N.j$ updates its color in response to a reset initiated by $j$. Distance 3 neighborhood of $j$ have stopped transmitting in response to the reset of $j$. Therefore, we can conclude that sensors in distance 2 neighborhood of $k$ have stopped transmitting. Now, when $k$ sends change color message $cm_k$, distance 1 neighbors of $k$ receive it successfully. When $j$ sends restart message, some distance 2 neighbors of $k$ are updated about the state of $k$. However, it is possible that when distance 1 neighbors of $k$ forward this restart, collisions may prevent some distance 2 neighbors of $k$ to not receive the update. On the other hand, $k$ will also witness this collision, and, as a result, schedules a reset computation in future frames. Hence, eventually, state of $k$ will be updated at all the sensors in its distance 2 neighborhood. □

We note that in this algorithm at most one neighbor is recovered in any reset. Therefore, if two or more sensors are involved in a collision at $j$ then $j$ still observes collisions after reset. Subsequent resets at $j$ or at other sensors will eventually restore collision-free communication at $j$. Thus, we have

**Theorem 3.4** *Eventually, the network self-stabilizes to the states where collision-free communication among the sensors is restored.* □

### 3.3 Illustration

In this section, we illustrate the TDMA slot assignment algorithm with an example. We consider the topology shown in Figure 8(a). The color assignments of each sensor is specified along with the ID of the node. For example, 2(1) denotes that sensor 2 is assigned color 1. Initially, we assume that $f_{current} = 0$ at all sensors. Based on initial color assignments, we can note that every sensor observes a collision. Sensors shown as filled circles denote that they have observed a collision.

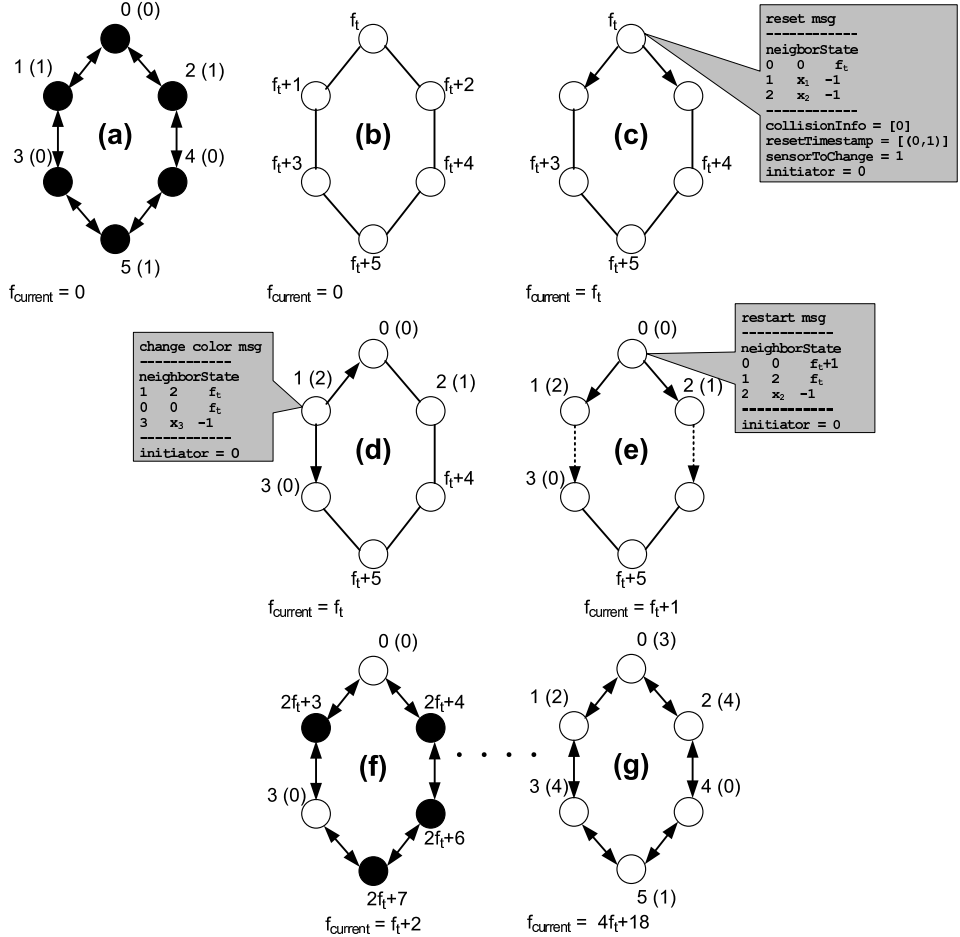

Figure 8: Illustration of the TDMA slot assignment algorithm

Assuming that the sensors have observed the collision for a threshold number of times, they schedule reset computation. Specifically, each sensor determines the frame in which it can send a reset message. Each sensor, say $j$, determines the frame for reset as follows: $f_{reset}.j = f_{current} + ID.j + f_t$, where $f_t = D3_{timeout}$. Figure 8(b) shows the frames in which the sensors have scheduled the reset computation.

In this illustration, as shown in Figure 8(c), sensor 0 sets $rm_0.sensorToChange = 1$. As a result, sensor 1 changes its color to 2. Then, it sends a change color message, $cm_1$ (cf. Figure 8(d)). Once sensor 0 receives $cm_1$, it updates its state and sends restart message, $sm_0$ (cf. Figure 8(e)). Once sensors 1 and 2 receive $sm_0$, they restart their communication. Continuing in this fashion, distance 3 neighborhood of sensor 0 restart communication. However, as we can observe from Figure 8(f), message communication is still not collision free. Sensors then schedule subsequent reset computations and, finally, as shown in Figure 8(g), collision-free communication is restored.

The convergence time for the network shown in Figure 8(a) is $4f_t + 18$ frames. If we had used the approach proposed in [10], where a base station initiates slot revalidation then the whole network is stopped (in response to missing token circulation). By contrast, in our algorithm, only the sensors in the distance 3 neighborhood of the initiator are stopped.

## 3.4 Convergence and Scalability

As discussed in Section 3.2, at most one sensor is recovered at the initiator in any reset computation. Assuming no failure of sensors, if a sensor observes $x$ collisions then, in the worst case, it takes $x$ resets to recover its neighbors. Additionally, we also note that reset computations are scheduled based on the IDs of the sensors in order to avoid interference among them. If two sensors that are not in the distance 3 neighborhood of each other initiate a reset then it is possible that resets are scheduled far apart even though there is no interference among them.

To improve the convergence and scalability of the algorithm, we can use the *neighborhood unique naming* scheme proposed in [8] that assigns unique IDs for sensors within any distance 3 neighborhood. This reduces the ID space of the network. Hence, two sensors not in the distance 3 neighborhood of each other can schedule their reset computations close to each other. As a result, the convergence time of the algorithm is improved.

Regarding scalability, first, we note that unlike [10], convergence can takes place in parallel if the initiators of resets are not in the distance 3 neighborhood of each other. And, since we adopt neighborhood unique naming scheme, the ID space of the network is small. Thus, integrating neighborhood unique naming scheme from [8] improves the convergence time and scalability of our algorithm.

## 3.5 Extensions

In this section, we discuss some extensions to our TDMA algorithm. First, in Section 3.5.1, we provide an extension that allows the sensors to deal with failure of their neighbors. Then, in Section 3.5.2, we present an approach to deal with addition of new sensors in the network. Subsequently, in Section 3.5.3, we present an optimization that that improves the bandwidth allocation of the sensors. Finally, in Section 3.5.4, we discuss how time synchronization can be achieved.

### 3.5.1 Dealing with Failure of Neighbors

In our algorithm, whenever a sensor (say $j$) hears a collision, it schedules a reset computation to restore collision-free communication. On the other hand, if $j$ does not hear a message or observe a collision in a given slot, it could be because of the one of the following factors: (i) suppose $k \in N.j$ is the neighbor that is assigned the corresponding color; $k$ may be failed, (iii) $k$ may have stopped in response to a reset, or (iii) $k$ does not have any data to send. If a sensor fails, the TDMA slots assigned to other sensors are still collision-free and, hence, normal operation of the network is not affected. However, the slots assigned to the failed sensors are wasted. In this section, we discuss an approach to reclaim slots assigned to failed sensors.

First, we introduce control message. Each sensor transmits a control message once in every $T_{control}$ frames. This message includes the color assignment of the sensor and its *nbrClr*. And, $T_{control}$ is determined when the network is deployed and is chosen based on how frequently the network changes. If topology changes are common, a smaller $T_{control}$ lets the sensors to quickly learn the state of their neighbors. On the other hand, a larger $T_{control}$ is more appropriate for a network that changes only occasionally.

To reclaim the slots assigned to failed sensors, we proceed as follows. Sensor $j$ concludes that $k \in N.j$ has failed if $f_{current}.j - lastReceived_k > T_{control}$, where $(k, -, lastReceived_k) \in nbrClr.j$. Specifically, if $j$ sees that it did not receive any message from $k$ for more than $T_{control}$ frames, it concludes that $k$ has failed.

When $j$ concludes $k$ has failed, it sets $(k, -, failed)$ in $nbrClr.j$. And, sends control message, $control_j$. Whenever a sensor receives $control_j$ such that $(k, -, failed) \in control_j.neighborState$, it marks $k$ as failed. The active neighbors of $j$ remove $(k, -, -)$ from *nbrClr* or *dist2Clr*. This allows the sensors to reuse the color assigned to $k$ to other sensors (in case of dynamic addition of new sensors or during reset computations). However, if $k$ has not failed, it announces its presences in its current TDMA slots by sending $control_k$. When neighbors of $k$ receive this message they update their $nbrClr$. Subsequently, distance 2 neighbors of $k$ also restore the state of $k$.

### 3.5.2 Dealing with Addition of Sensors

In this section, we discuss an approach to dynamically add new sensors in the network. This approach is similar to [10]. Suppose a sensor (say $p$) is added to the network such that the maximum degree of the network is not changed. Before $p$ starts transmitting application messages, it listens to the message communication of its neighbors. To let $p$ learn the colors used in its distance 2 neighborhood, we extend our algorithm as follows.

Sensor $p$ waits for $T_{control}$ frames before it participates in the network. This allows $p$ to learn distance 1 and distance 2 neighbors and their color assignments (from control messages of its neighbors). After $T_{control}$, $p$ chooses a color. Next, $p$ announces its presence to its neighbors by sending a control message in its newly computed slot. When a sensor receives a control message from $p$, it adds $p$ to its neighbor list and updates $nbrClr$. Subsequently, distance 2 neighbors of $p$ also learn its presence and update their $dist2Clr$.

Thus, this approach allows the addition of new sensors in a neighborhood such that it does not violate the maximum degree assumption. However, if two or more sensors are added simultaneously, it is possible that they may choose the same color and, as a result, cause collisions. Since our algorithm is self-stabilizing, the network will eventually self-stabilize to states where the color assignments are collision-free.

### 3.5.3 Improving the Bandwidth Allocation

In this section, we discuss an approach to improve the bandwidth allocation of the sensors. This approach allows the sensors to reduce the TDMA period and, hence, get better bandwidth allocation. The basic intuition behind this extension is that if $c_x$ is the maximum color used in the network, the ideal TDMA period should be $c_x + 1$.

In this approach, each sensor (say $j$) maintains $maxColor.j$ that denotes the maximum color used in its distance 2 neighborhood. It also maintains $controlMax.j$ that denotes the maximum color used in the network. Note that $j$ may not yet have the accurate information about the maximum color used in the network.

To improve the bandwidth allocation of the sensors, we extend the control message (discussed in Section 3.5.2) as shown in Figure 9. Any sensor in the network may decide to improve bandwidth allocation in the network. Let $j$ decides to improve bandwidth allocation. It sends a control message, $control_j$ that includes $control_j.maxColorInfo=\max(controlMax.j, maxColor.j)$. Sensor $j$ also indicates when the sensors can switch to new TDMA period, i.e., $control_j.switchOn = f_{switchOn}.j$, where $f_{switchOn}.j \geq f_{current}.j + 2T_{control}$. (We discuss why this is necessary later.)

| $control_j.maxColorInfo$ | $c = max(controlMax.j, maxColor.j)$, where $maxColor.j$ is the maximum in $\{c_x|(-,c_x,-) \in nbrClr.j \lor (-,c_x,-) \in dist2Clr.j \lor c_x = color.j\}$ |
|---|---|
| $control_j.switchOn$ | $f_{switchOn}.j$ |

Figure 9: Extending control message for improving bandwidth allocation

Whenever $k$ receives $control_j$ with $maxColorInfo$, $k$ sets $controlMax.k = \max(controlMax.k, control_j.maxColorInfo)$. It also notes down the frame in which it can switch to the new TDMA period, i.e., $f_{switchOn}.k = max(f_{switchOn}.k, control_j.switchOn)$. Thus, continuing in this fashion, each sensor will eventually learn the maximum color used in the network, i.e., *controlMax*. And, each sensor also knows the ideal TDMA period (i.e., $controlMax + 1$).

Once the sensors have learned the maximum color used in the network, they can update their TDMA period. However, this operation should occur synchronously. In other words, all the sensors should update their TDMA period at the same time. Otherwise, collisions may occur. To address this issue, first, we note the following. If the TDMA slots assigned to the sensors are consistent then all the sensors learn the maximum color used in the network in at most $2T_{control}$ frames, where $T_{control}$ is the period between two successive control messages (cf. Section 3.5.2). Since the initiator of this operation includes the frame in which new TDMA period is effective, each sensor knows exactly when to switch. Thus, the TDMA period can be updated to reflect the ideal value.

Additionally, a sensor can request for unused bandwidth in its distance 2 neighborhood using the approach proposed in [10] that negotiates with the neighbors to *lease* unused slots.

### 3.5.4 Time Synchronization

In our algorithm, we assume that the sensors have identical clocks. In this section, we show how our algorithm can be extended to synchronize time across the network. Again, in this extension, we use control message to synchronize time. Specifically, whenever a sensor (say $j$) transmits control message, $control_j$, it includes the information shown in Figure 10.

| $control_j.timesynchInfo$ | real time | current frame |
|---|---|---|
| | $wallClock.j$ | $f_{current}.j$ |

Figure 10: Extending control message for time synchronization

Specifically, the control message includes the real clock value at $j$ and current frame at $j$. Based on this information and the color assigned to $j$, a sensor determines start of the frame in wall-clock time. Whenever a neighbor (say $k \in N.j$) receives the control message, it updates $wallClock.k$ with $control_j.timesynchInfo$, if required. And, $k$ determines the start of the frame and updates it frame number and slot number accordingly.

Continuing in this fashion, time synchronization is achieved in the network. In addition, in the case where the TDMA slots are consistent, we can integrate synchronization algorithms proposed in the literature for sensor networks. For example, time synchronization services such as [16–18] could be integrated with our algorithm. These services synchronize the sensors to within a few microseconds. We also expect that performance of the time synchronization service will improve as TDMA provides a collision-free communication medium.

## 4 Related Work

Related work that deals with self-stabilizing deterministic slot assignment algorithms include [10, 11, 19, 20]. In [11], Kulkarni and Arumugam proposed self-stabilizing TDMA (SS-TDMA). In this algorithm, the topology of the network is known upfront and remains static. Also, a base station is responsible for periodic diffusing computations that assign/revalidate the slots of the sensors. Unlike [11], in the proposed algorithm, the sensors are aware of only their neighbors and there is no designated sensor that is responsible for the slot assignment.

In [10], Arumugam and Kulkarni proposed self-stabilizing deterministic TDMA algorithm. Again, this algorithm assumes the presence of the base station that is responsible for token circulation. And, the slots are assigned in sequential fashion. By contrast, the algorithm proposed in this paper assigns slots in a concurrent fashion and no token circulation is required.

In [19], Danturi et al proposed a self-stabilizing solution to dining philosophers problem where a process cannot share the critical section (CS) with non-neighboring processes also. This problem has application in distance-k coloring, where $k$ is the distance up to which a process cannot share CS. This algorithm requires each process $p$ to maintain a tree rooted at itself that spans the processes with whom $p$ cannot share CS using algorithms in the literature. However, the algorithm is written in shared memory model. On the other hand, the proposed algorithm can be used in sensor networks directly and also allows one to transform abstract programs into programs in WAC model.

In [20], BitMAC is proposed for collision-free communication in sensor networks. One of the important assumptions in this paper is that when two messages collide the result is an OR operation between them. This algorithm is not self-stabilizing. Unlike [20], our algorithm is written for WAC model and is self-stabilizing.

Related work that deals with randomized algorithms for TDMA slot assignment include [8, 12]. In [8], Herman and Tixeuil proposed a probabilistic fast clustering technique for TDMA slot assignment. In this algorithm, first, a maximal independent set that identifies the leaders is computed. These leaders are then responsible for distance 2 coloring. In [12], Busch et al proposed a randomized algorithm for slot assignment. The algorithm operates in two phases: (1) to compute the slots and (2) to determine the ideal TDMA period. Both these phases are self-stabilizing and can be interleaved. Unlike [8, 12], our algorithm is deterministic.

Several other TDMA algorithms are proposed in the literature (e.g., [21–24]). However, most of these algorithms are not self-stabilizing and/or make assumptions that are not relevant to WAC model or sensor networks.

## 5 Conclusion

In this paper, we presented a self-stabilizing deterministic TDMA slot assignment algorithm for write all with collision (WAC) model. We showed that the algorithm allows sensors to concurrently recover and self-stabilize starting from arbitrary states. Towards this end, we used distributed reset to restore the state of the network quickly. More specifically, whenever a sensor observes a collision for a threshold number of consecutive TDMA frames, it schedules a reset computation to recover its distance 2 neighborhood. We showed that at most one reset operation executes in a distance 3 neighborhood at any instant. And, we showed how the sensors recover in response to resets and reach legitimate states that satisfy the problem statement of TDMA.

Additionally, as discussed in [9], our algorithm is applicable in transforming existing programs in shared memory model into programs in WAC model. This allows one to reuse existing solutions in distributed computing for problems such as routing, data dissemination, synchronization, and leader election in the context of sensor networks. Thus, the algorithm proposed in this paper allows one to transform such solutions and evaluate them in sensor networks. As a result, we can rapidly prototype sensor network applications. (We refer the reader to [25] for examples of such transformations.) Moreover, this algorithm demonstrates the feasibility of a deterministic transformation of a program in shared memory model into a program in WAC model while preserving the self-stabilization property of the original program.

There are several possible future directions. While this algorithm demonstrates concurrent recovery of TDMA slots and the recovery time is expected to be reasonable for typical deployments, one future direction is to extend this algorithm to provide faster convergence. Additionally, while our experience in transformation of abstract programs to WAC model (cf. [25]) suggests that the efficiency of the program obtained by transformation is close to that of manually designed programs, another future direction is to quantify the efficiency of the transformed programs.